\documentclass[prd,superscriptaddress,amsfonts,amssymb,amsmath,showpacs,twocolumn]{revtex4-2}
\usepackage{bm}
\usepackage{amsfonts}
\usepackage{latexsym}
\usepackage[latin1]{inputenc}
\usepackage{graphicx}
\usepackage{amsmath}
\usepackage{palatino}
\usepackage{mathpazo}
\usepackage{textcomp}
\usepackage{float}
\usepackage{booktabs}
\usepackage{dcolumn}
\usepackage{hyperref}
\hypersetup{colorlinks,citecolor=blue}
\usepackage{amsmath}
\usepackage{xcolor}
\usepackage{orcidlink}
\usepackage[caption=false]{subfig}
\usepackage{commath}
\usepackage[greek, english]{babel}
\usepackage{alphabeta}
\usepackage{textgreek}
\newcommand{\be}{\begin{equation}}
\newcommand{\ee}{\end{equation}}
\newcommand{\ber}{\begin{eqnarray}}
\newcommand{\eer}{\end{eqnarray}}
\newcommand{\bern}{\begin{eqnarray*}}
\newcommand{\eern}{\end{eqnarray*}}
\newcommand{\beast}{\begin{equation*}}
\newcommand{\eeast}{\end{equation*}}
\newcommand{\nn}{\nonumber\\}

\newcommand{\tlnb}{\tilde{\nabla}^{}}

\def \D {\tilde{\nabla}}
\def \Th {\Theta}

\def\K{\mathcal{K}}

\def\lcdm{$\Lambda$CDM }
\def\LL{$\Lambda$ }
\def\bi{\begin{itemize}}
\def\ei{\end{itemize}}
\def\itt{\item}
\def\jnl@style{\it}
\def\aaref@jnl#1{{\jnl@style#1}}

\def\aaref@jnl#1{{\jnl@style#1}}

\def\aj{\aaref@jnl{AJ}}                   
\def\apj{\aaref@jnl{ApJ}}                 
\def\apjl{\aaref@jnl{ApJ}}                
\def\apjs{\aaref@jnl{ApJS}}               
\def\apss{\aaref@jnl{Ap\&SS}}             
\def\aap{\aaref@jnl{A\&A}}                
\def\aapr{\aaref@jnl{A\&A~Rev.}}          
\def\aaps{\aaref@jnl{A\&AS}}              
\def\mnras{\aaref@jnl{Mon.~Not.~Roy.~Astron.~Soc.}}             
\def\prd{\aaref@jnl{Phys.~Rev.~D}}        
\def\prc{\aaref@jnl{Phys.~Rev.~C}}  
\def\prl{\aaref@jnl{Phys.~Rev.~Lett.}}    
\def\qjras{\aaref@jnl{QJRAS}}             
\def\skytel{\aaref@jnl{S\&T}}             
\def\ssr{\aaref@jnl{Space~Sci.~Rev.}}     
\def\zap{\aaref@jnl{ZAp}}                 
\def\nat{\aaref@jnl{Nature}}              
\def\aplett{\aaref@jnl{Astrophys.~Lett.}} 
\def\apspr{\aaref@jnl{Astrophys.~Space~Phys.~Res.}} 
\def\physrep{\aaref@jnl{Phys.~Rep.}}      
\def\physscr{\aaref@jnl{Phys.~Scr}}       
\def\commat{\aaref@jnl{Comm.~Math.~Phys.}}              
\def\science{\aaref@jnl{Science}}               
\def\cqg{\aaref@jnl{Classical Quant.~Grav.}}            
\def\jpcs{\aaref@jnl{JPCS}}                                     
\def\ijmpd{\aaref@jnl{Int.~J.~Mod.~Phys.~D}}                    
\def\grg{\aaref@jnl{Gen.~Relat.~Gravit.}}               
\def\rpp{\aaref@jnl{Rep.~Prog.~Phys.}}          
\def\npa{\aaref@jnl{Nucl.~Phys.~A}}        
\def\lrr{\aaref@jnl{Living Rev.~Rel.}}                   
\def\jcap{\aaref@jnl{J.~Cosmology Astropart.~Phys.}}    
\def\rmp{\aaref@jnl{Rev.~Mod.~Phys.}}   
\def\epjc{\aaref@jnl{Eur.~Phys.~J.~C}}

\def\bw{\begin{widetext}}
\def\ew{\end{widetext}}

\allowdisplaybreaks[1]

\begin{document}

\color{black}       

\title{Perturbations in Bianchi Type-$I$ spacetimes with varying $\Lambda$, $G$}

\author{Alnadhief H. A. Alfedeel\orcidlink{0000-0002-8036-268X}}%
\email[Email:]{aaalnadhief@imamu.edu.sa}
\affiliation{Department of Mathematics and Statistics, Imam Mohammad Ibn Saud Islamic University (IMSIU),\\
Riyadh 13318, Saudi Arabia.}
\affiliation{Department of Physics, Faculty of Science, University of Khartoum, P.O. Box 321, Khartoum 11115, Sudan.}
\affiliation{Centre for Space Research, North-West University, Potchefstroom 2520, South Africa.}

\author{Maye Elmardi}
\email[Email: ]{maye.Elmardi@gmail.com}
\affiliation{Centre for Space Research, North-West University, Potchefstroom 2520, South Africa.}

\author{Amare Abebe\orcidlink{0000-0001-5475-2919}}
\email[Email: ]{amare.abebe@nithecs.ac.za}
\affiliation{Centre for Space Research, North-West University, Potchefstroom 2520, South Africa.}
\affiliation{National Institute for Theoretical and Computational Sciences (NITheCS), 3201 Stellenbosch, South Africa}

%
\date{\today}

\begin{abstract} 
In this paper, we investigate the evolution of cosmological perturbations within the context of Bianchi Type$-I$ spacetimes. We consider models containing viscous fluids with evolving cosmological (\LL) and Newtonian gravitational (G) parameters. The investigation of how over-densities in the viscous matter content in the Bianchi Type$-I$ model is our primary emphasis. In particular, we investigate the generation and propagation of signals associated with large-scale structures in this setting. We contrast our findings with the predictions of the classic $\Lambda$CDM ($\Lambda$-Cold-Dark-Matter) cosmological model to draw relevant contrasts and insights.

Our findings emphasise the need to incorporate viscous fluids into the Bianchi Type$-I$ geometry, as well as the dynamic fluctuations of \LL and $G$. These factors influence the rate of structure growth in the cosmos as a whole. Thus, our findings offer light on the complex dynamic interaction between viscosity, changing cosmological parameters, and the growth of large-scale structures in an anisotropic universe.

\end{abstract}

\maketitle

\date{\today}

\section{Introduction}
In the field of general relativity, Bianchi Type$-I$ spacetimes are a class of cosmological models that describe homogeneous and anisotropic universes. These spacetimes are characterized by a set of three independent scale factors that determine the expansion rate of the universe in different directions. Perturbations in such spacetimes refer to small deviations from the homogeneous and isotropic behavior, which can have significant implications for the evolution of the universe.

One interesting aspect of Bianchi Type$-I$ spacetimes is the possibility of considering variations in the cosmological constants, namely the cosmological constant $\Lambda$ and the gravitational constant $G$. The cosmological constant $\Lambda$ is related to the energy density of the vacuum, while the gravitational constant $G$ determines the strength of the gravitational force. Varying these constants can lead to changes in the overall dynamics of the universe and affect the evolution of perturbations and hence the large-scale structure- growth rate.

Recent cosmological observations \cite{perlmutter1997measurements,perlmutter1998discovery,perlmutter1999measurements,riess1998observational} have fundamentally reshaped our understanding of the universe, compelling the scientific community to grapple with the concept of late-time cosmic acceleration. Within the framework of the standard $\Lambda$-cold dark matter ($\Lambda$CDM) model, this cosmic acceleration is ascribed as dark energy, purportedly constituting approximately 70\% of the cosmic matter-energy content. The remaining 25\% is attributed to cold dark matter, a form of noninteracting matter detected solely through its gravitational effects, existing primarily in the electromagnetic spectrum.

One seemingly straightforward explanation for dark energy is the incorporation of the cosmological constant $\Lambda$ into the field equations, endowing it with the role of a vacuum energy source. However, this apparent simplicity belies the intricate challenges associated with the cosmological constant, including the cosmological constant problem and the coincidence problem. Consequently, alternative theories and approaches have been explored, encompassing modifications to the theory of gravity itself, the introduction of spacetime inhomogeneities and/or anisotropies, deviations from the standard Friedman-Lema\^{i}tre-Robertson-Walker (FLRW) spacetime model, and the consideration of additional matter properties such as viscosity.

Furthermore, a compelling avenue of exploration involves relaxing the assumption of the constancy of both $\Lambda$ and $G$, the universal gravitational constant, which feature prominently in the cosmological field equations. While the FLRW model assumes homogeneity and isotropy on large cosmic scales, the current state of cosmological knowledge, wherein nearly 95\% of the universe's fluid components remain enigmatic, necessitates a comprehensive exploration of unconventional paradigms. Hence, our motivation for scrutinizing one of the Bianchi spacetime models, which deviate from the standard cosmological model by preserving homogeneity while introducing non-isotropic characteristics.

Intriguingly, according to the Dirac hypothesis \cite{Dirac1937} concerning the evolution of fundamental constants, the possibility arises to contemplate the time-varying nature of $\Lambda$ and $G$ and their far-reaching cosmological consequences \cite{Alfedeel:2018thd,alfedeel2020bianchi1,Vishwakarma1996a,Vishwakarma1996b,Vishwakarma1999,Vishwakarma2000,Vishwakarma2001,Vishwakarma2005,bali2012bianchi, singh2010bulk,padmanabhan1987viscous}. In this context, the study of perturbations in Bianchi Type$-I$ spacetimes with varying $\Lambda$ and $G$ has attracted considerable attention in recent years. Researchers have investigated the effects of such variations on the growth of perturbations, the formation of structures, and the overall cosmological evolution. These studies have important implications for our understanding of the universe and can provide valuable insights into the nature of dark energy, which is thought to be responsible for the observed acceleration of the universe's expansion.

Alfedeel and Abebe \cite{alfedeel2022evolution} conducted an investigation into the cosmological implications of time-dependent cosmological $\Lambda$ and Newtonian $G$ ``constants" within the framework of Bianchi Type$-I$ spacetime. They employed known cosmological data, such as $\Omega_m,\Omega_r$, and $\Omega_\Lambda$, to solve Einstein's field equations. Furthermore, Alfedeel \cite{alfedeel2020bianchi2} explored a Bianchi Type$-I$ cosmological model characterized by homogeneity and anisotropy, featuring time-varying Newtonian and cosmological constants. These investigations provided analytical solutions for the average scale factor, incorporating hypergeometric functions.

In this study, we investigate the cosmological perturbations within an isotropic Bianchi Type$-I$ cosmological model, incorporating an imperfect viscous fluid and time-variable Newtonian and cosmological constants, along similar lines of the recent work \cite{aade} for Bianchi Type-V spacetimes. Notably, our approach avoids assuming precise formulas for $\Lambda$ or $G$, in contrast to other approaches. Instead, the system of differential equations naturally self-consolidates from its definitions. Our primary focus centers on the evolution of over-densities in the Bianchi$-I$ model, leading to the production of large-scale structures, with comparative analysis against traditional $\Lambda$CDM solutions.

The subsequent sections of this work are structured as follows: Section~\ref{efes} provides a concise overview of the Bianchi Type-$I$ background cosmological model. Perturbations around this Bianchi Type-$I$ background are elucidated in Section~\ref{pert}. Section~\ref{resdis} presents various cosmological models contingent upon the selection of time-varying shear and bulk viscosities and discuss the outstanding results. Ultimately, we conclude our article in Section~\ref{conc}, summarizing our findings and implications.

\section{Background field equations}\label{efes}

The Bianchi type$-V$ line-element in orthogonal space and time coordinates is represented
by the following formula:
\begin{equation}
ds^2=dt^2-A^{2}dx^2- B^{2}dy^2+C^{2}dz^2~.\label{metric}
\end{equation}
where $A=A(t)$, $B=B(t)$ and $C=C(t)$ are the metric potential and $m$ is constant.
We assume that the universe is filled by a viscous fluid whose distribution
in space is represented by the following energy momentum tensor \cite{bds,tasahs}:
\begin{align}
T_{ij} & = (\rho  + \overline{p} ) u_{i} u_{j} + \overline{p} g_{ij}-2\eta \sigma_{ij} ~,\label{Tij} \\
\sigma_{ij} & =    \nabla_k u_{i} \; h_j^k + \nabla_k \dot{u}_{j} \; h_i^k - \frac{1}{3} \theta h_{ij}~,\nonumber \\
h_{ij} & = g_{ij} + u_i u_j\;,\nonumber \\
\overline{p} &= p - \left( \xi -\frac{2}{3} \eta \right) \nabla_iu_i = p - \left( 3 \xi - 2 \eta \right) H\;.
\end{align}
Here $\rho$ is matter energy density, $p$ is the isotropic pressure, $u^i$ is
$4$-velocity vector of the cosmic fluid and it is time-like quantity that satisfies $ u_i u^i = -1$, $\sigma_{ij}$ is the shear
and $\overline{p}$ is the effective pressure, $\xi$ and $\eta$ stand for the bulk
and shear viscosity coefficients respectiively and $h_{ij}$ is the projection tensor.
The gravitational field equations of Einstein with time-varying $G$ and $\Lambda$ in geometrical units where $ c = 1$ are given by
\begin{equation}
R_{ij} - \frac{1}{2} g _{ij} R = - \kappa G T_{ij}+ \Lambda g_{ij}\;.\label{EFEs}
\end{equation}
Here  $\kappa\equiv 8\pi$ and $R_{ij}$ is Ricci tensor, $R$ is Ricci scalar and $g_{ij}$ is the symmetric
second-rank metric tensor. Using Eqs. \eqref{metric}-\eqref{Tij}, the 
EFEs in \eqref{EFEs} are calculated as follows:
\begin{align}
\frac{\ddot{B}}{B} +  \frac{\ddot{C}}{C} +  \frac{\dot{B}}{B}\frac{\dot{C}}{C} - 2\kappa G \eta \frac{\dot{A}}{A} & =  -\kappa G p  + \kappa G \left( \xi -\frac{2}{3} \eta \right) + \Lambda~,\label{Gtt}\\
\frac{\ddot{A}}{A} +  \frac{\ddot{C}}{C} +  \frac{\dot{A}}{A}\frac{\dot{C}}{C} - 2\kappa G \eta \frac{\dot{B}}{B} & =  -\kappa G p  + \kappa G\left( \xi -\frac{2}{3} \eta \right) + \Lambda~,\label{Gxx}\\
\frac{\ddot{A}}{A} +  \frac{\ddot{B}}{B} + \frac{\dot{A}}{A}\frac{\dot{B}}{B}  - 2 \kappa G \eta \frac{\dot{C}}{C}  & = -\kappa G p + \kappa G \left( \xi -\frac{2}{3} \eta \right) + \Lambda~,\label{Gyy}\\
\frac{\dot{A}}{A}\frac{\dot{B}}{B} + \frac{\dot{A}}{A}\frac{\dot{C}}{C} + \frac{\dot{B}}{B}\frac{\dot{C}}{C} & = \kappa G \rho + \Lambda~,\label{Gzz}\\
\end{align}
where an overdot represents partial differentiation with respect to cosmic time $t$. On the other hand, the
vanishing of the Einstein tensor in Eq. \eqref{EFEs} and energy momentum tensor $\nabla^j T_{ij}=0$ in
Eq. \eqref{Tij} produces an auxiliary equation as  

\bw\begin{align}
\kappa G\left[ \dot{\rho} + (\overline{p} + \rho)\left( \frac{\dot{A}}{A} + \frac{\dot{B}}{B}  
 + \frac{\dot{C}}{C}  \right)\right ] + \kappa \rho \dot{G} + \dot{\Lambda}
 - 4\kappa G \eta \sigma^2=0\;.\label{ConsEq} 
\end{align}\ew
This equation can be split into tow equations
\begin{eqnarray}
&&\dot{\rho} + 3 H \left[p + \rho - (3\xi - 2\eta) H\right]- 4 \eta \sigma^2 =  0~, \label{Rho+Evol}\\
&&\kappa \rho \dot{G}  + \dot{\Lambda} = 0  \;,\label{GL+Evol}
\end{eqnarray}
thus showing the co-evolution of the gravitational and cosmological parameters. It is particularly interesting to note that, according to Eq. \eqref{GL+Evol},  a monotonically increasing/decreasing $G$ necessarily implies a decreasing / increasing \LL provided that over time, $\rho$ remains positive.

For the Bianchi type$-I$ spacetime the average scale factor $a$, and the Hubble parameter $H$, expansion scalar $\theta$, shear
scalar $\sigma$, and deceleration parameter $q$ are defined as 
\bw\begin{align}
a^3 & = ABC =V~, \nonumber\\ 
H &= \frac{\dot{a}}{a} = \frac{1}{3}\left(H_{x}+H_{y}+H_{z}\right) = \frac{1}{3} \left( \frac{\dot{A}}{A} + \frac{\dot{B}}{B} + \frac{\dot{C}}{C}  \right)~,\nonumber \\
\sigma^2 &= \frac{1}{2}\sigma_{ij}\sigma^{ij} = \frac{1}{6}\left[ \left( \frac{\dot{A}}{A}
-\frac{\dot{B}}{B}\right) ^{2}
+\left( \frac{\dot{B}}{B}-\frac{\dot{C}}{C}\right) ^{2}
+\left( \frac{\dot{C}}{C}-\frac{\dot{A}}{A}\right)^{2}\right] = \frac{\sigma_0^2}{a^6}~,\label{sigmaEq}\nonumber \\
q        &= -\frac{a\ddot{a}}{\dot{a}^{2}}=-\frac{\dot{H}}{H^{2}}-1\;,
\end{align}\ew
where $\sigma_0$ is a constant that is related to the universe anisotropy and where $ H_{x}\;, H_{y}$ and $H_{z}$ are the directional Hubble parameters
along $x,y$ and $z$ directions respectively. 
The average anisotropy parameter $A_p$ is defined as
\begin{equation}
A_p   = \frac{1}{3}\sum_{i=1}^{3} \left(\frac{H_i - H}{H} \right)^2\;.
\end{equation}
The field equations \eqref{Gxx}-\eqref{Gzz} can be integrated to give 
\begin{eqnarray}
&&\frac{\dot{A}}{A}-\frac{\dot{B}}{B} = \frac{k_1}{a^3} e^{-2\kappa \int G \eta  dt} ~,\label{AB}\\
&&\frac{\dot{B}}{B}-\frac{\dot{C}}{C} = \frac{k_2}{a^3} e^{-2\kappa \int G \eta dt} ~,\label{BC}\\
&&\frac{\dot{A}}{A}-\frac{\dot{C}}{C} = \frac{k_3}{a^3} e^{-2\kappa \int G \eta dt} ~,\label{AC}
\end{eqnarray}
$k_1$, $k_2$ and $k_3$ are constant of integration.  Integrating these equations one more time gives an
expression for the metric functions $A$, $B$ and $C$ as 
\begin{align} 
A &= A_0 \;a\; \exp \left\{ \frac{k_1+k_2}{3} \int \frac{e^{-2\kappa \int G \eta dt}}{a^3} ~dt\right\} ~,\label{Aeq}\\
B &=  B_0 \;a\; \exp \left\{ \frac{k_2-k_1}{3} \int \frac{ e^{-2\kappa \int G \eta  dt} }{a^3} ~dt\right\}~,\label{Beq}\\
C &= C_0 \;a\; \exp\left\{ \frac{-2k_2-k_1}{3} \int \frac{ e^{-2\kappa \int G \eta  dt}  }{a^3} ~dt\right\}~,\label{Ceq} 
\end{align}
where $A_0$, $B_0$, and $C_0$ are constants of integration satisfying  the following relations:
\bw
\begin{eqnarray}
A_0  = \sqrt[3]{k_1 k_2} \;, \quad B_0 = \sqrt[3]{k_1^{-1}k_3}\;,\quad C_0=\sqrt[3]{(k_1 k_3)^{-1}}\;,\quad  A_0B_0C_0=1~.
\end{eqnarray}
\ew
Using the values of $A, B$ and $C$, Eq.\eqref{sigmaEq} gives a dierect expression for $\sigma$ in terms of bulk viscosity $\eta$ as:
\begin{equation}
 \sigma = \frac{\sigma_0}{V^2} e^{-2\kappa \int  G \eta  dt}\;.
\end{equation}
The generalized Friedmann equations for Bianchi type-$I$ spacetimes with a viscous-fluid are achieved 
from \eqref{Gtt}-\eqref{Gzz} as
\begin{eqnarray}
&&\kappa G \overline{p}-\Lambda = H^{2}(2q-1)-\sigma^{2} ~,\label{Friedman1}\\
&&\kappa G \rho +\Lambda =3H^{2}-\sigma^{2} \;.\label{Friedman2}
\end{eqnarray}
The generalized Raychaudhuri equation reads:
\begin{align}
\dot{H} + 3H^2  - \Lambda + \frac{\kappa G }{2}(p - \rho) - \frac{3}{2} \kappa G \left( \xi - \frac{2}{3} \eta\right)H =0\;.\label{dHdtEq}
\end{align}
We note that the evolution of the universe in the Bianchi type$-I$ cosmological model with time-dependent $G$ and $\Lambda$ term for a viscous fluid is governed by variables $\rho, G, \Lambda, H$, $A$, $B$ and $C$, but
Eqs. \eqref{Rho+Evol}, \eqref{GL+Evol}, \eqref{dHdtEq} and \eqref{Aeq}-\eqref{Ceq} provide only
6 equations. In order to close the system of these equations a supplementary equation is required. Thus, 
we can divide and re-arrange the Friedmann equation \eqref{Friedman2} by $3H^2$ as 
\begin{equation}
\label{constr}
1 =  \Omega_m + \Omega_\Lambda + \Omega_\sigma\;. 
\end{equation}
Now, differentiating Eq. \eqref{constr} with respect to time will provide an extra the extra evolution equation:
\begin{align}
\dot{\Omega}_\Lambda & = - \dot{\Omega}_\sigma  - \dot{\Omega}_m\;.   \label{dOldt}
\end{align}
Here we have used  the standard definitions of fractional energy densities:
\begin{align}
\Omega_{m}\equiv \frac{\kappa G \rho_{m}}{3H^2}~, \qquad
\Omega_{ \Lambda} \equiv \frac{\kappa G \rho_{\Lambda }} {3H^2}~,\qquad 
\Omega_{\sigma}\equiv \frac{\sigma^2}{ 3H^2}~, \qquad
\end{align}
with their present-day values given by
\begin{align}
 \Omega_{m_0}= \frac{\kappa G_0 \rho_{m_0}}{3H^2_0}~, \qquad
\Omega_{ \Lambda_0} = \frac{\kappa G_0 \rho_{\Lambda_{0} }} {3H^2_0}\;,\qquad 
\Omega_{\sigma_0}= \frac{\sigma^2_{0} }{ 3H^2_0}\;.
\end{align}

We can then rewrite  Eqs. \eqref{Rho+Evol} and \eqref{GL+Evol}as: 
\bw\begin{align}\label{omdot}
\dot{\Omega}_m & = - \left\{ 2\frac{\dot{H}}{H} - \frac{\dot{G}}{G} \right\} \Omega_m  - H (3+3w_m)\Omega_m + 3 \kappa G \left( \xi - \frac{2}{3} \eta \right) + 4 \eta \kappa G  \Omega_\sigma\;,\\
\dot{G}  & = - \frac{G}{\Omega_m}\; \dot{\Omega}_\Lambda   -  2\frac{\dot{H}}{H} \frac{\Omega_\Lambda}{\Omega_m} G \;,\label{Ogdot}\\
\dot{\Omega}_\sigma & = - \left(6H + 2\frac{\dot{H}}{H}\right)\Omega_\sigma\;.\label{Osdot}
\end{align}\ew
Introducing dimensionless parameters corresponding to $H$ and $G$ will further simplify our computations, so we define:
\be
 h\equiv \frac{H}{H_0}\;, \qquad \mathcal{G} \equiv \frac{G}{G_0}\;.
\ee
The bulk and shear viscosities can be parametrised as  \cite{aade}:
\be
\qquad \xi=  \alpha H_0(\rho_{\rm m}/\rho_{\rm m0})^n~, \qquad \mbox{and} ~~~\eta = \beta H\;,
\ee
where the coefficients  $\alpha$, $\beta$, and  the exponent $n$ are dimensionless constants of the viscosities involved, with $0 \leq n \leq \frac{1}{2}$ \cite{pavon1991causal,maartens1995dissipative,zimdahl1996bulk,santos1985isotropic}.

\begin{figure}
 \includegraphics[scale=0.65]{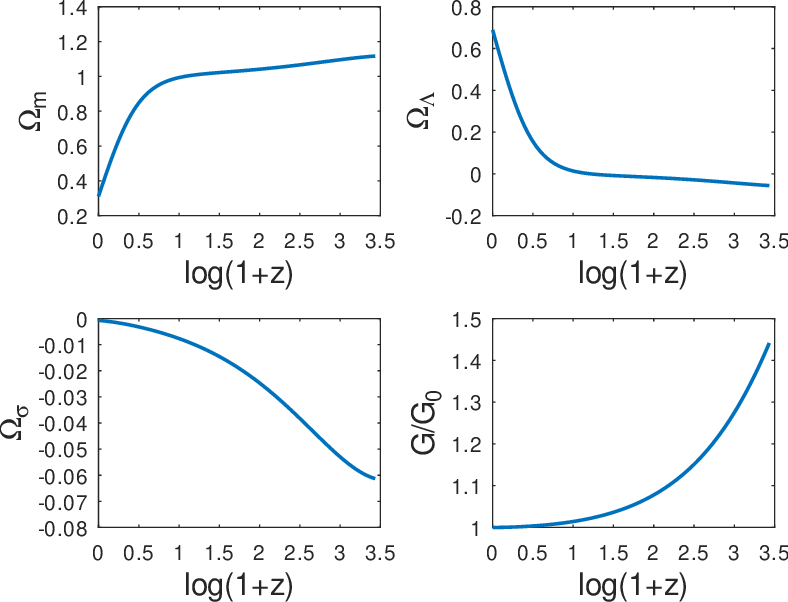}
\caption{This figures show the variation of viscous Bianchi type-I  characterized parameters
 $[\Omega_m, \Omega_\Lambda, \Omega_\sigma, h, \mathcal{G}]$ with redshift. The initial conditions
 $\Omega_m(0) =0.3111, \Omega_\Lambda(0) = 0.6889$, $\Omega_\sigma(0) = 1-\Omega_m(0) - \Omega_\Lambda(0)$,  $h(0)=1, \mathcal{G}(0)=1$ are used 
 alongside the Runge-Kutta method to integrate the background differential equations numerically.}
 \label{fig1}
\end{figure}

We can now rewrite a fully dimensionless system of equations in redshift space using the relation
$\dot{Q}= -(1+z)HQ'$ for any time-dependent quantity $Q$,  thus re-expressing  Eqs.  \eqref{dHdtEq},  \eqref{dOldt}, \eqref{omdot}, \eqref{Ogdot}, and \eqref{Osdot} as follows:
\bw\begin{align}
h' &= \frac{h}{(1+z)} \left[ 3 - 3\Omega_\Lambda -  \frac{3}{2} (1-w_m) \Omega_m \right]  - \frac{3}{2} \frac{\kappa G_0 }{(1+z)} \left[  \alpha \left( \frac{ h^2 \Omega_m}{g \Omega_{m_0}} \right)^n - \frac{2}{3} \beta h  \right] \mathcal{G} \;,   \label{dhdzDE} \\     
\Omega'_m  & =   - \left\{ \frac{2h'}{h}- \frac{\mathcal{G}'}{\mathcal{G}} \right\} \Omega_m  + \frac{\left(3 + 3w_m\right)}{1+z}  \Omega_m  - \frac{3}{h} \frac{\kappa G_0 }{(1+z)}
 \left[  \alpha \left( \frac{h^2 \Omega_m }{ \mathcal{G} \Omega_{m_0}}\right)^n - \frac{2}{3} \beta h \right]\;\mathcal{G} - \frac{ 4\beta \kappa G_0}{(1+z)} \Omega_{\sigma} \mathcal{G} \; ,
               \label{dOmdzDE}\\
\mathcal{G}' &= - \frac{\Omega'_\Lambda}{\Omega_m}\;\mathcal{G}   -  2\frac{h'}{h} \frac{\Omega_\Lambda}{\Omega_m} \mathcal{G}\;,\label{OgdzDE}\\
\Omega'_\sigma&= \frac{2}{1+z}  \left[ 3\Omega_\Lambda +  \frac{3}{2} (1-w_m) \Omega_m \right] \Omega_\sigma   +  \frac{3\kappa G_0 }{(1+z)h} \left[  \alpha \left( \frac{ h^2 \Omega_m}{g \Omega_{m_0}} \right)^n - \frac{2}{3} \beta h  \right] \;\mathcal{G}\; \Omega_\sigma\;,\\
\Omega'_\Lambda&  = \left \{ \frac{2h'}{h}- \frac{\mathcal{G}'}{\mathcal{G}} \right\} \Omega_m  
               -  \frac{ \left(3 + 3w_m\right)}{1+z}  \Omega_m  
               -  \frac{3\kappa G_0(1-\Omega_\sigma) }{(1+z)h}  \left[  \alpha \left( \frac{h^2 \Omega_m }{ \mathcal{G} \Omega_{m_0}}\right)^n - \frac{2}{3} \beta h \right]\; \mathcal{G} \nonumber\\
               & + \frac{ 4\beta \kappa G_0}{(1+z)} \Omega_{\sigma} \;\mathcal{G}
               - \frac{2}{1+z}  \left[ 3\Omega_\Lambda +  \frac{3}{2} (1-w_m) \Omega_m \right] \Omega_\sigma\;.\label{olevz}
\end{align}\ew
The above equations are first-order coupled differential equations that describe the
evolution of $h$, $\Omega_m$, $\mathcal{G}$, $\Omega_\Lambda$ and $\Omega_\sigma$
with respect to the redshift $z$. It is interesting to note that we did not assume any mathematical formula for $G$ and $\Lambda$, we only let them smoothly and naturally appear from the manipulations of field equations. As depicted in plots of Fig. \ref{fig1}, we numerically solve these equations using latest observational results as the initial conditions for the matter and dark energy components, and the values of $h$ and $\mathcal{G}$ both normalised to unity today, by definition. The plots clearly show the expected transition from 

\section{Perturbations}\label{pert}
Despite the assumption of isotropy and homogeneity on the largest possible cosmological scales, the real universe is lumpy. There are galaxies, clusters, voids, filaments, and walls if one zooms on to smaller patches of the observable universe. These structures are believed to have been seeded in the early universe due to gravitational instabilities, and amplified through some mechanism that we can describe by the use of cosmological perturbation theory.  The standard way of dealing with the perturbations involves the metric perturbation theory \cite{lifshitz46, bardeen80, KS1984} whereas we follow the covariant approach of perturbations \cite{ehlers61,hawking66, olson76,eb89,dbe92,bde92,clarkson2003} as no unphysical gauge modes appear in this approach (see~\cite{gidelew} and the references therein for more details).

In the $1+3$ covarriant formalism, we usually start the analysis by defining the covariant and gauge-invariant gradient variables that describe perturbations in the matter energy density,  expansion and shear \cite{eb89,carloni, aadd, aade}:

\be
D_{a}\equiv\frac{a\tlnb_{a}\rho}{\rho}\;,\quad  Z_{a}\equiv a\tlnb_{a}\Th\;,\quad  \Sigma_{a}\equiv a\tlnb_{a}\sigma\;.
\ee
These gradient variables evolve according to the following equations:
\bw\ber
&&\label{pertd}\dot{D}_a  - \left[w-\left(\xi-\frac{2\beta}{9}\Theta\right)\frac{\Th}{\rho}  +  
\frac{\left( n\xi \Th- w \rho \right)(4\beta \sigma^2/3\rho) }{(1+w)\rho - \left(\xi-\frac{2\beta}{9}\Theta\right)\Th }   
- \frac{4\beta \sigma^2}{3\rho} \right]\Th D_a -  \sigma_a^b D_b \nonumber \\
          &&\quad+\left[1+w-\left(\xi-\frac{2\beta}{9}\Theta\right)\frac{\Th}{\rho}
          + \frac{ \left( \xi-\frac{4\beta}{9}\Theta \right) (4\beta \sigma^2/3) }{(1+w)\rho - \left(\xi-\frac{2\beta}{9}\Theta\right)\Th }  \frac{\Th}{\rho} 
- \frac{4\beta \sigma^2}{3\rho} \right]Z_a
+ \frac{8\beta}{3} \frac{\Th}{\rho} \sigma  \Sigma_a \nonumber \\
          &&\quad+  \left[ 1- \frac{\frac{4\beta \sigma^2}{3}}{ (1+w)\rho - \left(\xi-\frac{2\beta}{9}\Th\right)\Th}\right] \; \frac{2\beta \Th}{3 \rho}\; \sigma_a^b Z_b =0\;, \\
&&\label{pertz}\dot{Z}_a  +\left[\frac{2}{3}\Th-\frac{3\kappa G}{2}\left(\xi-\frac{4\beta}{9}\Th\right)-\frac{\left(\xi - 4\beta/9\Theta \right)\dot\Th}{(1+w)\rho - \left(\xi - 2\beta/9 \Theta \right)\Theta}\right]Z_a-\frac{\left(\xi - 4\beta/9\Theta \right)}{\rho+p - \left(\xi - 2\beta/9 \Theta \right)\Theta}\tlnb{^2} Z_a \nonumber \\
&&\quad+\left[\frac{\kappa G}{2}(1+3w)\rho-\frac{3\kappa G}{2}n\xi\Th-\frac{\left(n\xi\Theta-w\rho \right)\dot{\Th}}{(1+w)\rho - \left(\xi - 2\beta/9 \Theta \right)\Theta}\right]D_a-\frac{\left(n\xi\Theta-w\rho \right)}{\rho+p - \left(\xi - 2\beta/9 \Theta \right)\Theta}\tlnb{^2} D_a \nonumber \\
&&\quad+ 4\sigma \Sigma_a + \left[ 1 - \frac{ (2\beta/3)\dot\Th }{\rho+p - \left(\xi - 2\beta/9 \Theta \right)\Theta} \right]  \; \sigma_a^b Z_b 
- \frac{ 2\beta/3 }{\rho+p - \left(\xi - 2\beta/9 \Theta \right)\Theta} \tlnb{^2} \; \sigma_a^b Z_b =0\;, \\
&&\label{perts}\dot{\Sigma}_a + \Th\Sigma_a + \sigma \left[1 +  \frac{ \left(\xi-\frac{4\beta}{9}\Theta\right) \Th}{(1+w)\rho - \left(\xi-\frac{2\beta}{9}\Theta\right)\Th } \right]Z_a  
                 + \sigma \left[ \frac{\left( n\xi \Th- w \rho \right)}{(1+w)\rho - \left(\xi-\frac{2\beta}{9}\Th\right)\Th} \right] \Th D_a \nonumber \\
               && \quad+ \frac{\frac{2\beta}{3} \sigma \Th }{ (1+w)\rho - \left(\xi-\frac{2\beta}{9}\Theta\right)\Th } \; \sigma_a^b Z_b - \sigma^b_a \Sigma_b =0\;.
\eer\ew
These are the equations governing the rate at which structures grow in a Bianchi type$-I$ universe with viscous matter and varying $\Lambda$ and $G$ terms as previously prescribed.
We must integrate these equations in parallel with background evolution equations Eqs. \eqref{dhdzDE}- \eqref{olevz} and analyse the results.  In order to be able to solve the system, we follow similar arguments as in \cite{aade} and ignore  terms like where $\sigma_a^b Z_b$, $ \sigma_a^b D_b$ and $\sigma^b_a \Sigma_b$  introduced due to the effect of shear. The observed anisotropy in the universe is expected to be very small though not necessarily negligible, so any product of the shear and a first-order perturbed quantity should be even smaller, and hence negligible compared to the other terms. Such an assumption will lead to the more simplified system of equations below:

\bw\ber
&&\label{pertd}\dot{D}_a  - 
\left[w-\left(\xi-\frac{2\beta}{9}\Theta\right)\frac{\Th}{\rho}  +  
\frac{\left( n\xi \Th- w \rho \right)(4\beta \sigma^2/3\rho) }{(1+w)\rho - \left(\xi-\frac{2\beta}{9}\Theta\right)\Th }   
- \frac{4\beta \sigma^2}{3\rho} \right]\Th D_a \nonumber \\
          &&\quad+\left[1+w-\left(\xi-\frac{2\beta}{9}\Theta\right)\frac{\Th}{\rho}
          + \frac{ \left( \xi-\frac{4\beta}{9}\Theta \right) (4\beta \sigma^2/3) }{(1+w)\rho - \left(\xi-\frac{2\beta}{9}\Theta\right)\Th }  \frac{\Th}{\rho} 
- \frac{4\beta \sigma^2}{3\rho} \right]Z_a
+ \frac{8\beta\Th}{3\rho} \sigma  \Sigma_a=0\;,\\
&&\label{pertz}\dot{Z}_a  +\left[\frac{2}{3}\Th-\frac{3\kappa G}{2}\left(\xi-\frac{4\beta}{9}\Th\right)-\frac{\left(\xi - 4\beta/9\Theta \right)\dot\Th}{(1+w)\rho - \left(\xi - 2\beta/9 \Theta \right)\Theta}\right]Z_a-\frac{\left(\xi - 4\beta/9\Theta \right)}{\rho+p - \left(\xi - 2\beta/9 \Theta \right)\Theta}\tlnb{^2} Z_a \nonumber \\
&&\quad+\left[\frac{\kappa G}{2}(1+3w)\rho-\frac{3\kappa G}{2}n\xi\Th-\frac{\left(n\xi\Theta-w\rho \right)\dot{\Th}}{(1+w)\rho - \left(\xi - 2\beta/9 \Theta \right)\Theta}\right]D_a\nn
&&\quad-\frac{\left(n\xi\Theta-w\rho \right)}{\rho+p - \left(\xi - 2\beta/9 \Theta \right)\Theta}\tlnb{^2} D_a+ 4\sigma \Sigma_a=0 \;,\\
&&\label{perts}\dot{\Sigma}_a + \Th\Sigma_a + \sigma \left[1 +  \frac{ \left(\xi-\frac{4\beta}{9}\Theta\right) \Th}{(1+w)\rho - \left(\xi-\frac{2\beta}{9}\Theta\right)\Th } \right]Z_a  
                 + \sigma \left[ \frac{\left( n\xi \Th- w \rho \right)}{(1+w)\rho - \left(\xi-\frac{2\beta}{9}\Th\right)\Th} \right] \Th D_a=0\;.
\eer\ew

We will now extract the scalar components of the perturbations as most of large-scale structure formation is believed to come through spherical clustering (for which the scalar components are responsible):

\be
\Delta\equiv a\tlnb{^a}D_a\;, \quad Z\equiv a\tlnb{^a}Z_a\;,\quad \Sigma\equiv a\tlnb{^a}\Sigma_a\;.
\ee
The evolution equations in these scalar variables are then given by:

\bw\ber
 &&\label{perds}\dot{\Delta}-
 \left[w-\left(\xi-\frac{2\beta}{9}\Theta\right)\frac{\Th}{\rho}  +  
 \frac{\left( n\xi \Th- w \rho \right)(4\beta \sigma^2/3\rho) }{(1+w)\rho - \left(\xi-\frac{2\beta}{9}\Theta\right)\Th }   
 - \frac{4\beta \sigma^2}{3\rho} \right]\Th \Delta \nonumber \\
           &&\quad+\left[1+w-\left(\xi-\frac{2\beta}{9}\Theta\right)\frac{\Th}{\rho}
           + \frac{ \left( \xi-\frac{4\beta}{9}\Theta \right) (4\beta \sigma^2/3) }{(1+w)\rho - \left(\xi-\frac{2\beta}{9}\Theta\right)\Th }  \frac{\Th}{\rho} 
 - \frac{4\beta \sigma^2}{3\rho} \right] Z\nn
 && \quad+ \frac{8\beta\Th}{3\rho} \sigma  \Sigma=0\;,\\
  &&\label{pertzs}
  \dot{Z}+\left[\frac{2}{3}\Th-\frac{3\kappa G}{2}\left(\xi-\frac{4\beta}{9}\Th\right)-\frac{\left(\xi - 4\beta/9\Theta \right)}{(1+w)\rho - \left(\xi - 2\beta/9 \Theta \right)\Theta}(\dot\Th+\frac{2\K}{a^2})\right]Z\nn
 &&\quad-\frac{\left(\xi - 4\beta/9\Theta \right)}{(1+w)\rho - \left(\xi - 2\beta/9 \Theta \right)\Theta}\tlnb{^2} Z\nn
  &&\quad
  +\left[\frac{\kappa G}{2}(1+3w)\rho-\frac{3\kappa G}{2}n\xi\Th-\frac{\left(n\xi\Theta-w\rho \right)}{(1+w)\rho - \left(\xi - 2\beta/9 \Theta \right)\Theta}(\dot\Th+\frac{2\K}{a^2})\right]\Delta\;,\nn
  &&\quad-\frac{\left(n\xi\Theta-w\rho \right)}{(1+w)\rho - \left(\xi - 2\beta/9 \Theta \right)\Theta}\tlnb{^2} \Delta+ 4\sigma \Sigma=0\nn
  &&\label{pertss}\dot{\Sigma} + \Th\Sigma + \sigma \left[1 +  \frac{ \left(\xi-\frac{4\beta}{9}\Theta\right) \Th}{(1+w)\rho - \left(\xi-\frac{2\beta}{9}\Theta\right)\Th } \right]Z  
                   + \sigma \left[ \frac{n\xi \Th- w \rho}{(1+w)\rho - \left(\xi-\frac{2\beta}{9}\Th\right)\Th} \right] \Th \Delta=0\;,\nn
 \eer\ew
 where we have used the relation
\bw \be
 a\D^a\left(\D^2Z_a\right) =\D^2\left(a\D^aZ_a\right)  + \frac{2K}{a^2} a \D^a Z_a = \D^2Z+\frac{2\K}{a^2}Z\;.
 \ee\ew
In the above commutation relation ${\mathcal{K}}$ denotes the curvature scalar of the 3-space. It is worth noting that the evolution equations above are still partial differential equations, as the perturbations are both position and time dependent. Such equations are generally too complicated to solve, so we convert them to ordinary differential equations by using the decomposition assumption

\be
\D^2X=-\frac{k^2}{a^2}X
\ee
for some averaged wavenumber $k$ and then write down  the evolution of the perturbations in the $kth$ mode as follows:
\bw\ber
&&\label{perdh}\dot{\Delta}^k -
\left[w-\left(\xi-\frac{2\beta}{9}\Theta\right)\frac{\Th}{\rho}  +  
\frac{\left( n\xi \Th- w \rho \right)(4\beta \sigma^2/3\rho) }{(1+w)\rho - \left(\xi-\frac{2\beta}{9}\Theta\right)\Th } 
- \frac{4\beta \sigma^2}{3\rho} \right]\Th \Delta^k \nonumber \\
          &&\quad+\left[1+w-\left(\xi-\frac{2\beta}{9}\Theta\right)\frac{\Th}{\rho}
          + \frac{ \left( \xi-\frac{4\beta}{9}\Theta \right) (4\beta \sigma^2/3) }{(1+w)\rho - \left(\xi-\frac{2\beta}{9}\Theta\right)\Th }  \frac{\Th}{\rho} 
- \frac{4\beta \sigma^2}{3\rho} \right] Z^k + \frac{8\beta\Th}{3\rho} \sigma  \Sigma^k=0\;,\\
&&\label{pertzh}\dot{Z}^k+\left[\frac{2}{3}\Th-\frac{3\kappa G}{2}\left(\xi-\frac{4\beta}{9}\Th\right)-\frac{\left(\xi - 4\beta/9\Theta \right)}{(1+w)\rho - \left(\xi - 2\beta/9 \Theta \right)\Theta}(\dot\Th -\frac{k^2}{a^2})\right]Z^k\nn
&&\quad
+\left[\frac{\kappa G}{2}(1+3w)\rho-\frac{3\kappa G}{2}n\xi\Th-\frac{\left(n\xi\Theta-w\rho \right)}{(1+w)\rho - \left(\xi - 2\beta/9 \Theta \right)\Theta}(\dot\Th -\frac{k^2}{a^2})\right]\Delta^k+ 4\sigma \Sigma^k=0\;,\nn
&&\label{pertsh}\dot{\Sigma}^k + \Th\Sigma^k + \sigma \left[1 +  \frac{ \left(\xi-\frac{4\beta}{9}\Theta\right) \Th}{(1+w)\rho - \left(\xi-\frac{2\beta}{9}\Theta\right)\Th } \right]Z^k  
                 + \sigma \left[ \frac{n\xi \Th- w \rho}{(1+w)\rho - \left(\xi-\frac{2\beta}{9}\Th\right)\Th} \right] \Th \Delta^k=0\;.\nn
\eer\ew
By defining the following dimensionless quantities
\[  \gamma \equiv \frac{k^2}{H_0^2}\;, \qquad  \mathcal{Z} \equiv \frac{Z}{H_0}~,\qquad \mathcal{S} \equiv \frac{\Sigma}{H_0}\;, \] 
and expressing the differential equations in redshift space, we can rewrite the harmonically-decomposed perturbations equations as:

\bw\ber\label{hdsed}
{\Delta'}^k && = -\frac{3}{(1+z)} \Bigg\{ w -  \frac{\kappa G_0 \mathcal{G} }{ \Omega_{m} h} \left[ \alpha \left(\frac{h^2  \Omega_{m} }{ \mathcal{G} \Omega_{m0} }\right)^n - \frac{2\beta}{3} h \right]           
-
\frac{4 \beta \kappa G_0 \mathcal{G} }{3}   \frac{\Omega_{\sigma} }{ \Omega_{m} } 
\nonumber \\
&& \quad \quad + 
\frac{4\beta \kappa G_0 \mathcal{G}}{3} \frac{\Omega_\sigma}{ \Omega_{m}}   
\left( \frac{ \alpha n \left(\frac{h^2   \Omega_{m}}{ \mathcal{G}   \Omega_{m0} }\right)^n - \frac{ w h \Omega_{m}}{\kappa G_0 \mathcal{G} } }{  \frac{ (1+w) h \Omega_{m} }{\kappa G_0 \mathcal{G}} -  \left[ \alpha \left(\frac{h^2   \Omega_{m} }{ \mathcal{G}   \Omega_{m0} }\right)^n - \frac{2\beta}{3} h \right] }
  \right)~\Bigg\} \Delta^k\nn
&&\quad + \frac{1}{h(1+z)}
\Bigg[ 1+ w -  \frac{\kappa G_0 \mathcal{G}}{ \Omega_{m} h} \left[ \alpha \left(\frac{h^2 \Omega_{m} }{ \mathcal{G} \Omega_{m0} }\right)^n - \frac{2\beta}{3} h \right]           
 -
 \frac{ 4 \beta \kappa G_0  \mathcal{G} }{3} \frac{\Omega_{\sigma} }{ \Omega_{m} }
\nonumber \\
&& \quad \quad + 
\frac{4 \beta \kappa G_0 \mathcal{G}}{3} \frac{\Omega_\sigma}{\Omega_{m}}  \left( \frac{ \alpha \left(\frac{h^2 \Omega_{m}}{ \mathcal{G} \Omega_{m0} }\right)^n - \frac{ 4\beta}{3}h }{  \frac{ (1+w) h \Omega_{m} }{\kappa G_0 \mathcal{G}} -  \left[ \alpha \left(\frac{h^2 \Omega_{m}}{ \mathcal{G} \Omega_{m0} }\right)^n - \frac{2\beta}{3} h \right] }
  \right)~\Bigg]
\mathcal{Z}^k
\nonumber \\
 &&\quad \quad
+ \frac{ 8\beta}{3} \frac{\kappa G_0  \mathcal{G} }{h(1+z)} \frac{\sqrt{3 \Omega_\sigma}}{\Omega_m}\; \mathcal{S}^k
\eer
\ew

\bw\ber\label{hdsez}
\mathcal{Z}'^k && = \frac{1}{(1+z)\;h} \Bigg[ 2h - \frac{3\kappa G_0 \mathcal{G} }{2} \left\{ \alpha \left(\frac{h^2   \Omega_{m}}{ \mathcal{G}   \Omega_{m0} }\right)^n - \frac{4\beta}{3} h \right\} \nonumber \\ 
&& \qquad -\frac{  \alpha \left(\frac{h^2   \Omega_{m}}{ \mathcal{G}   \Omega_{m0} }\right)^n - \frac{4\beta}{3} h }{ \frac{(1+w) h^2   \Omega_{m} }{ \kappa G_0 \mathcal{G}} - h  \left[ \alpha \left(\frac{h^2   \Omega_{m}}{ \mathcal{G}   \Omega_{m0} }\right)^n - \frac{2\beta}{3} h \right] } 
\left( -h h'(1+z)  - \frac{\gamma}{3} (1+z)^2 \right) \Bigg]  \mathcal{Z}^k  \nonumber \\
&&\qquad + \frac{3}{(1+z) } \Bigg[ \frac{  \Omega_{m} h}{2} (1+3w) - \frac{3 \alpha n \kappa G_0 \mathcal{G}}{2} \left(\frac{h^2   \Omega_{m} }{\mathcal{G}   \Omega_{m0} }\right)^n   
\nonumber \\
&&\qquad -\frac{ \alpha n \left(\frac{h^2   \Omega_{m}}{ \mathcal{G}   \Omega_{m0} }\right)^n - \frac{ w h \Omega_{m} }{ \kappa G_0 \mathcal{G} } }{  
\frac{ (1+w) h^2 \Omega_{m}}{\kappa G_0 \mathcal{G} }  -  h \left\{ \alpha \left(\frac{h^2   \Omega_{m}}{ \mathcal{G}   \Omega_{m0} }\right)^n - \frac{2\beta}{3} h \right\} }  
\left( -h h'(1+z)  - \frac{\gamma}{3} (1+z)^2 \right)
\Bigg] \Delta^k
 \nn
 &&\qquad
+ \frac{4 \sqrt{  \Omega_\sigma}}{(1+z)} \mathcal{S}^k\;,
\eer\ew

\bw
\ber\label{hdses}
{\mathcal{S}'}^k && = \frac{3}{(1+z)} \mathcal{S}^k +
\frac{\sqrt{3  \Omega_\sigma}}{(1+z)} \left[1 + 
\frac{ \alpha \left(\frac{h^2   \Omega_{m} }{ \mathcal{G}   \Omega_{m0} }\right)^n - \frac{ 4\beta}{3}h }{  \frac{ (1+w) h \Omega_{m} }{\kappa G_0 \mathcal{G} } -  \left[ \alpha \left(\frac{h^2   \Omega_{m} }{ \mathcal{G}   \Omega_{m0} }\right)^n - \frac{2\beta}{3} h \right]}
\right] \mathcal{Z}^k  \nonumber \\
&& \quad\quad+ \frac{3 h \sqrt{3\Omega_\sigma}}{(1+z)} \left[  \frac{ \alpha n \left(\frac{h^2   \Omega_{m}}{ \mathcal{G}   \Omega_{m0} }\right)^n - \frac{ w h \Omega_{m}}{\kappa G_0 \mathcal{G} } }{  \frac{ (1+w) h \Omega_{m} }{\kappa G_0 \mathcal{G} } -  \left[ \alpha \left(\frac{h^2   \Omega_{m} }{ \mathcal{G}   \Omega_{m0} }\right)^n - \frac{2\beta}{3} h \right] }
   \right] \Delta^k\;.
\eer\ew

In the next section, we will set up the initial conditions for the perturbations and analyse the effect of the viscosity parameters on the growth of the perturbations.

\section{Results and Discussion}\label{resdis}

The harmonically decomposed scalar evolution equations given in Eqs.  \eqref{hdsed}-\eqref{hdses} form a closed system of ODEs that, given initial conditions, can be solved numerically.
We set set our initial conditions at some redshift $z_{in}$, and we plot the amplitudes of the perturbations normalised by the initial conditions:
\be
\delta^k(z)\equiv \frac{\Delta^k(z)}{\Delta^k(z_{in})}\;.
\ee
We use Planck 2018 results for the background cosmological parameters and set the initial conditions for the perturbations at $z_{in}=20$. In the following, we will make a comparative analysis of the results obtained in relation to the viscosity parameters as presented in Figs. \ref{fig2}, \ref{fig3}, \ref{fig4} and \ref{fig5}:

\bi
\itt Keeping all other factors the same, the perturbation amplitudes increase with increasing wavelength (decreasing $\gamma)$.
\itt In both the short wavelength and long wavelength limits, on a fixed wavelength scale, increasing $\beta$ increases the amplitude of the perturbations while increasing $\alpha$ decreases the amplitude. Also, the smaller the exponent $n$, the higher the amplitude.
\ei

The above results suggest that, as one should suspect, bulk viscosity suppresses late-time structure formation since it introduces resistance to matter clumping. Increasing the shear viscosity, on the other hand, encourages more clumping, and hence more structures to form.


\begin{figure}[H]
  \centering
\includegraphics[width=0.45\textwidth]{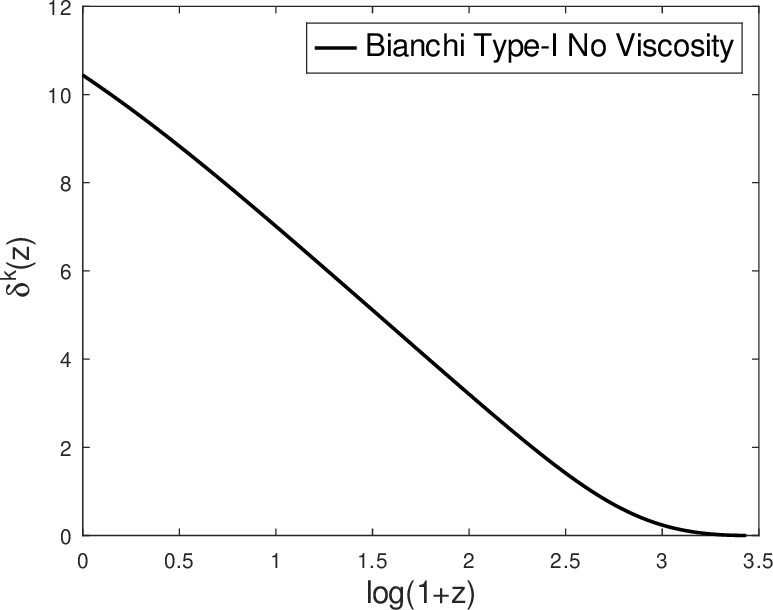}
\end{figure}

\bw
\begin{figure}[H]
  \centering
  \subfloat[Short wave length]{\includegraphics[width=0.5\textwidth]{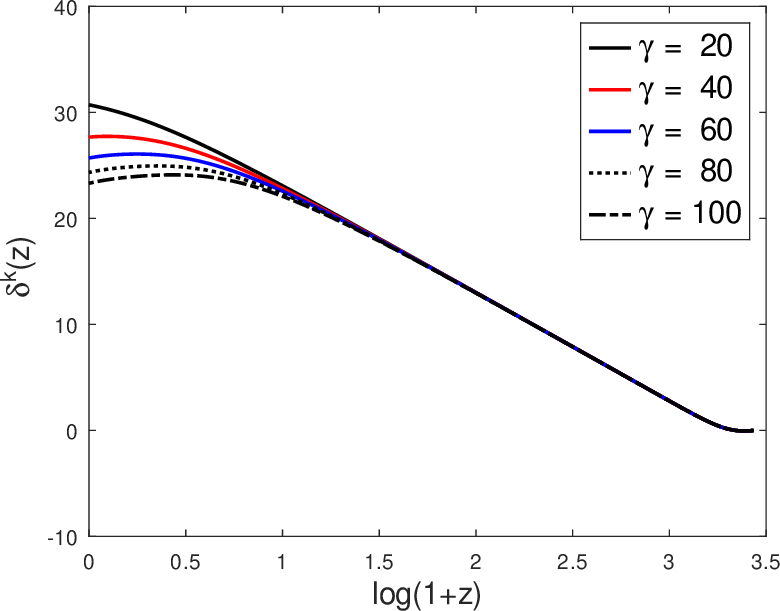}}
  \subfloat[Long wave length]{\includegraphics[width=0.5\textwidth]{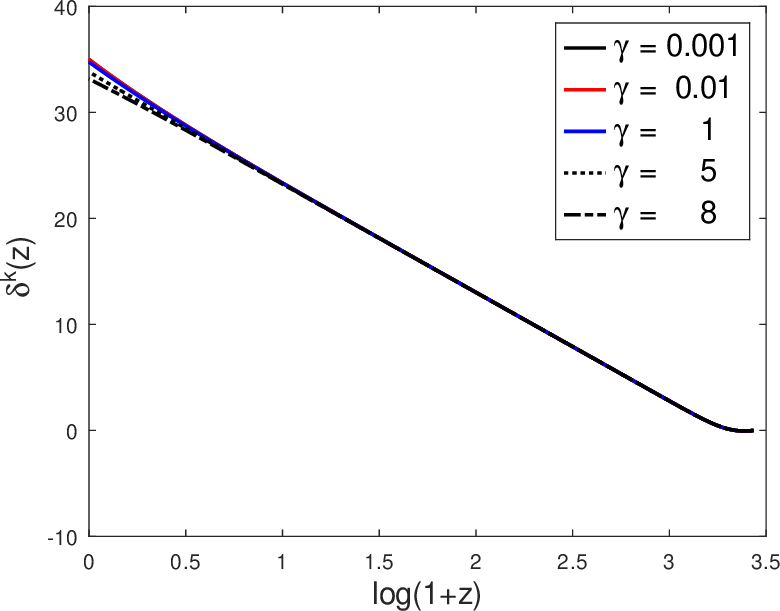}}
  \caption{The variation of the matter density perturbations $\delta^k(z)$ for viscous Bianchi type$-I$ cosmological model vs redshift.
  The initial conditions $\Delta^k (z_0) = 10^{-5}$, $\mathcal{Z}^k (z_0) = 10^{-5}$ and $\mathcal{S} (z_0) = 10^{-5}$  are used with $n=0.2$ to
  integrate the governing system of perturbations along with background evolution equations.}
      \label{fig2}
\end{figure}
\ew

\bw\begin{figure}[H]
  \centering
  \subfloat[Short wave length]{\includegraphics[width=0.5\textwidth]{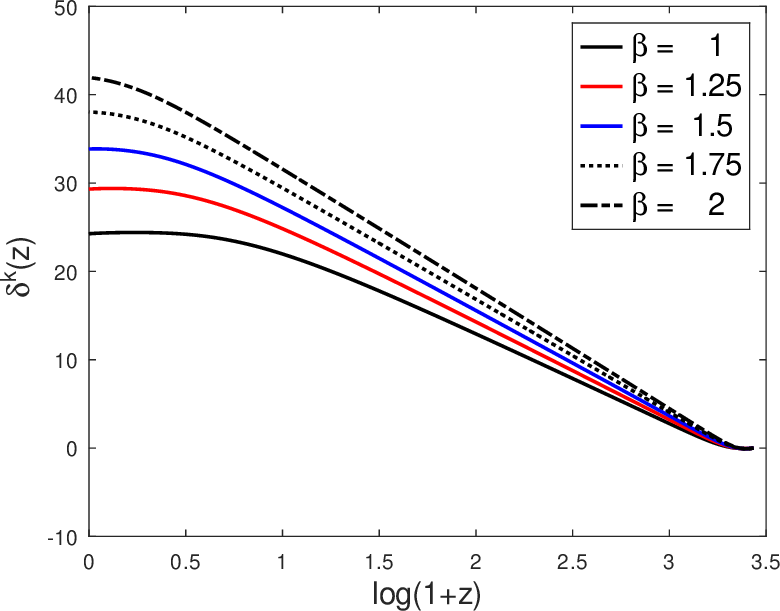}}
  \subfloat[Long wave length]{\includegraphics[width=0.5\textwidth]{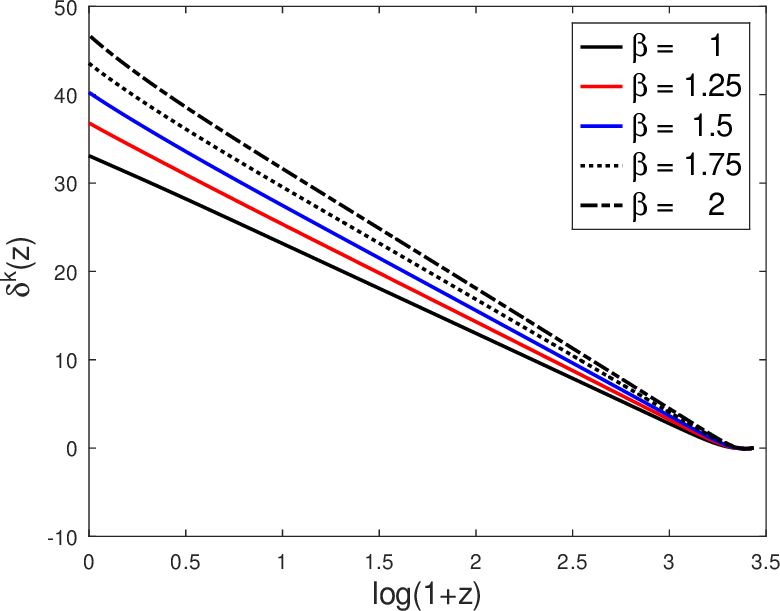}}
  \caption{The variation of the matter density perturbations $\delta^k(z)$ for viscous Bianchi type$-I$ cosmological model vs redshift $z$.
  The initial conditions $\Delta^k (z_0) = 10^{-5}$, $\mathcal{Z}^k (z_0) = 10^{-5}$ and $\mathcal{S} (z_0) = 10^{-5}$  are used with $n=0.2$, $\alpha=0.3$, $\gamma=4$ and $\gamma=40$ and different values of $\beta$.}
      \label{fig3}
\end{figure}
\ew


\bw\begin{figure}[H]
  \centering
  \subfloat[Short wave length]{\includegraphics[width=0.5\textwidth]{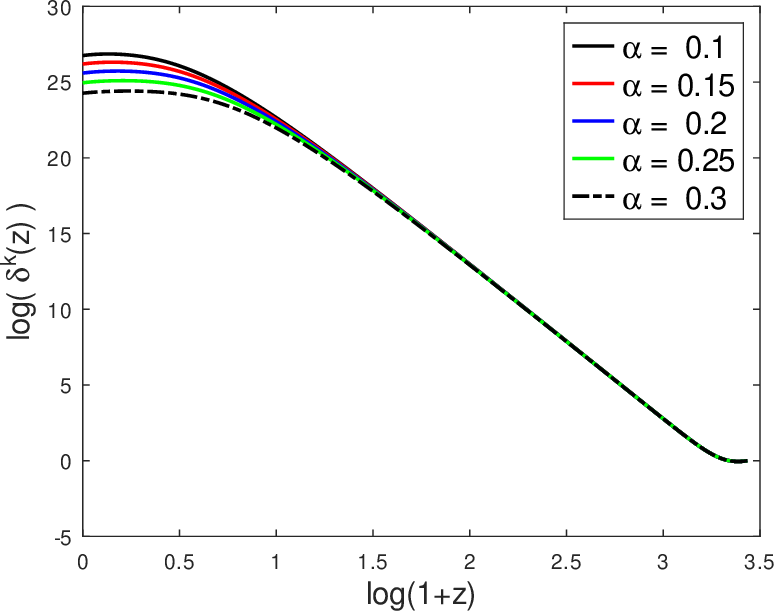}}
  \subfloat[Long wave length]{\includegraphics[width=0.5\textwidth]{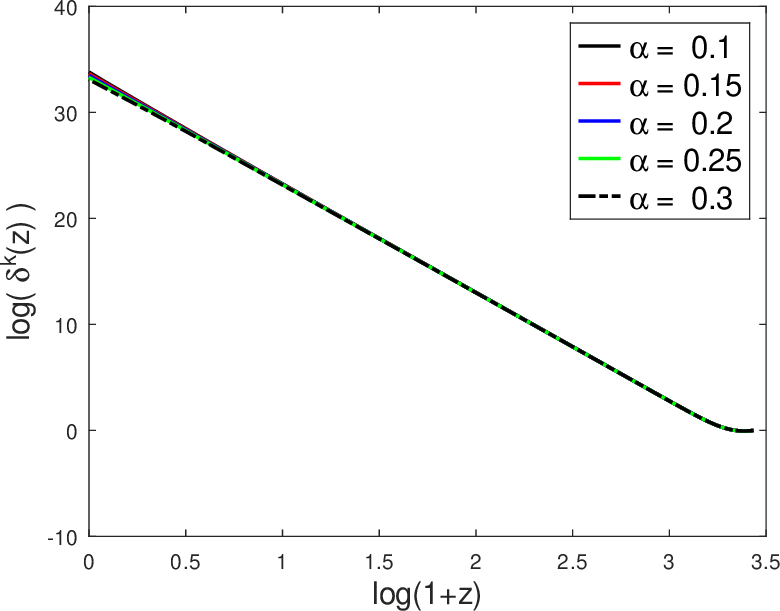}}
  \caption{The variation of the matter density perturbations $\delta^k(z)$ for viscous Bianchi type$-I$ cosmological model vs redshift $z$.
  The initial conditions $\Delta^k (z_0) = 10^{-5}$, $\mathcal{Z}^k (z_0) = 10^{-5}$ and $\mathcal{S} (z_0) = 10^{-5}$ are used with $n=0.2$, $\beta=1$, $\gamma=4$ and $\gamma=40$ and different values of $\alpha$.}
      \label{fig4}
\end{figure}
\ew

\bw\begin{figure}[H]
  \centering
  \subfloat[Short wave length]{\includegraphics[width=0.5\textwidth]{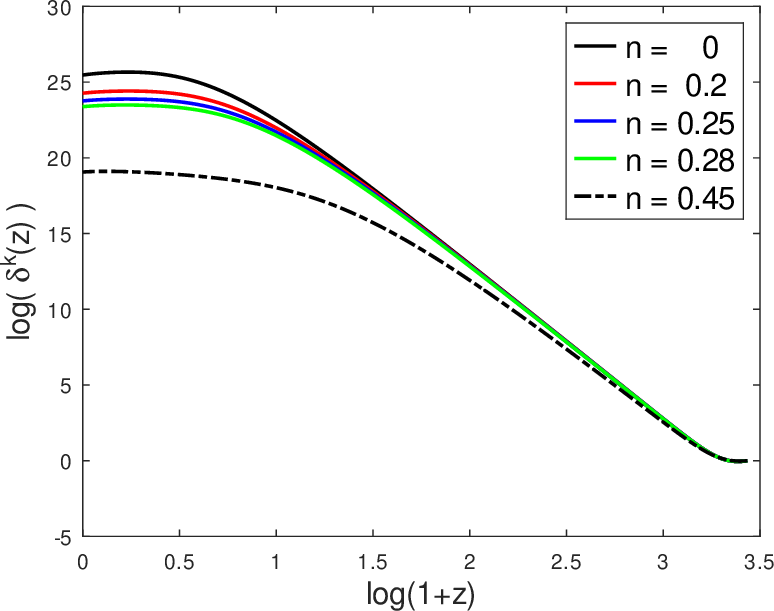}}
  \subfloat[Long wave length]{\includegraphics[width=0.5\textwidth]{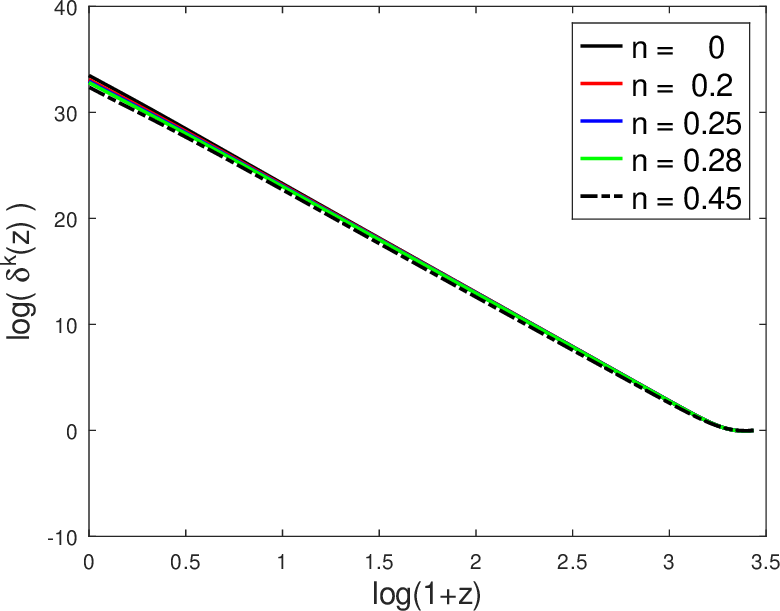}}
  \caption{The variation of the matter density perturbations $\delta^k(z)$ for viscous Bianchi type$-I$ cosmological model vs redshift $z$.
  The initial conditions $\Delta^k (z_0) = 10^{-5}$, $\mathcal{Z}^k (z_0) = 10^{-5}$ and $\mathcal{S} (z_0) = 10^{-5}$ are used with $\alpha=0.3$, $\beta=1$, $\gamma=4$ and $\gamma=40$ and different values of $n$.}
      \label{fig5}
\end{figure}

\ew

\section{Conclusion}\label{conc}

In this study, we explored the perturbations in the Bianchi type$-I$ cosmological model filled with an imperfect (viscous) fluid and evolving Newtonian and cosmological parameters, G and \LL  respectively. Our investigation aimed to understand the evolution of perturbations and their impact on the formation of large-scale structures, comparing our results to the traditional \lcdm model results.

We started by considering the background evolution of the universe, allowing for the variation of $G$ and \LL over time. This nonstandard cosmological model departs from the standard FLRW cosmology, permitting  anisotropies in spacetime and imperfections in the fluid distributions.  Then, using the $1 + 3$ covariant perturbation formalism, we derived a set of equations describing the evolution of perturbations in matter energy density, expansion, and shear. These perturbations are crucial for understanding the growth of large-scale structures in the universe. 

Our findings reveal that viscosity has a significant impact on the growth of large-scale structure. In particular, we showed that the bulk viscosity tends to suppress growth whereas shear viscosity enhances it, as depicted in Figs. \ref{fig1}-\ref{fig5}.

We compared our results to the traditional $\Lambda$CDM model, which assumes a constant cosmological constant and a Newtonian gravitational constant. Our findings highlight the differences and potential advantages of considering time-variable ``constants" and an imperfect viscous fluid in the cosmological model can impact the formation of large-scale structures in the universe. While further research and observations are needed to validate these findings, they offer a valuable contribution to our understanding of the cosmos beyond the standard cosmological paradigm.
\section{Acknowledgements}
The authors extend their appreciation to the Deputyship for Research \& Innovation, Ministry of Education in Saudi Arabia for funding this research through the project number IFP-IMSIU-2023122 . The authors also appreciate the Deanship of Scientific Research at Imam Mohammad Ibn Saud Islamic University (IMSIU) for supporting and supervising this project

\end{document}